# Network attack detection at flow level

Aleksey A. Galtsev [*] and Andrei M. Sukhov

Samara State Aerospace University, Moskovskoe sh., 34, Samara, 443086, Russia;
*e-mails:* `galaleksey@gmail.com`, `amskh@yandex.ru`

**Abstract.** In this paper, we propose a new method for detecting unauthorized network intrusions, based on a traffic flow model and Cisco NetFlow protocol application. The method developed allows us not only to detect the most common types of network attack (DDoS and port scanning), but also to make a list of trespassers' IP-addresses. Therefore, this method can be applied in intrusion detection systems, and in those systems which lock these IP-addresses.

## 1 Introduction

Currently, Internet information resources are actively growing, penetrating many spheres of social life. Information technologies are being introduced not only into private enterprises, but also in the provision of public services. With each passing day, more and more confidential transactions are carried out via the Internet. In connection with these trends, the question of computer networks security is starkly raised. Attackers have developed and actively use many types of network intrusion [1,2,3,4], most of which can be prevented by standard methods of protection.

This article focuses on detecting and preventing network attacks of two types that are impossible to prevent by the standard settings of information resource software. This are "Distributed Denial of Service" attacks (DDoS attacks) [4,5] and port scanning, that are used to find bottlenecks in network information systems. In recent years the rate of end-user connections to the Internet has increased sharply, which has given rise to an increase in the number and intensity of attacks such as DDoS. These attacks are highly damaging to the information service, and at the same time simple in their execution. Port scanning is used by hackers to conduct "network intelligence". In this article we would like to propose a new method of detecting DDoS attacks and port scanning, based on the Cisco NetFlow protocol [7,8].

NetFlow is a network protocol developed by Cisco Systems for collecting IP traffic information. NetFlow has become an industry standard for traffic monitoring and is supported by platforms other than Cisco IOS (Internetwork Operating System) and NXOS (Nexus Operating System) [9].

---

[*] corresponding author

A network flow has been defined in many ways. The traditional Cisco definition is to use a 7-tuple key, where a flow is defined as a unidirectional sequence of packets sharing all of the following 7 values:

- Source IP address
- Destination IP address
- Source port for UDP or TCP, 0 for other protocols
- Destination port for UDP or TCP, type and code for ICMP, or 0 for other protocols
- IP protocol
- Ingress interface (SNMP ifIndex)
- IP Type of Service

The proposed method for detecting network attacks based on the traffic flow model is described in [10]. Traffic models shows that two parameters, the load of the channel and the number of active flows in it, must be used for a full representation of the network state. In this paper, the criteria of abnormal network conditions, which can determine the start of the attack, were formulated. A more detailed model is described in Section 2. The values of these parameters can be measured using the NetFlow protocol, implemented on Cisco routers. In [11] and [12], the authors suggested that the traffic flow model can be used for network security problems, in particular to detect network attacks such as DDoS, port scanning and network worms. Also in [13] and [14], an attempt was made to use the NetFlow protocol to detect DoS attacks such as Smurf and worms W32.Blaster Worm and Red Worm.

The aim of this work is to show that the NetFlow protocol can be used for the detection of DDoS attacks and port scanning, and to formulate an algorithm to identify IP-addresses from which the attack is carried out. This algorithm enables to the creation of "black lists" of addresses that should be blocked to prevent the attack. This article is organised as follows:

- Section 2 - describes the flow traffic model, on which a method for detecting network attacks has been built
- Section 3 - experiment to study the various attacks
- Section 4 - the definition of the detection algorithm under consideration
- Section 5 - describes the Research Center of DDoS attacks at the Samara State Aerospace University

## 2  Traffic model

In this paper, we would like to propose a method of diagnosing the backbone links and testing it on existing networks. This method is based on a traffic model [10], according to which the number of active flows can be considered as an important characteristic of the real network state. Two variables, the number of active flows and the utilisation of the channel, best describe the current network state. Analysis of all data on the network, represented by individual points on the plane with axes, which are plotted the number of active flows and utilization of the network, allows to the definition of three areas that correspond to qualitatively different states of the network.

It has been previously shown [10] that the first part of the curve formed from average values of the data produces a straight line, which ends with an inflection point. The straight line corresponds to the part of the network that is characterised by a minimal loss of IP-packets, less than a half of one percent. The bent part of the curve corresponds to an overloaded network, and is characterised by a significant packet loss of up to 5%, which reduces the effective size of the transferred segment of the TCP/IP. The third, nearly horizontal, portion of the curve corresponds to a completely unusable network with significant packet loss of over 5%.

The distribution of total load tends to a normal (Gaussian) distribution, since the total load of the studied channel is the result of multiplexing a large number of flows that are independent of each other. The theoretical model allows us to estimate the confidence interval for the working area of the curve:

$$B(t) = b(N + kA(\epsilon)\sqrt{N}) \qquad (1)$$

Here $A(\epsilon)$ is the normal quantile function. Equation 1 indicates that the real state of the network, described by the number of active flows $N$ and the flow data rate $B(t)$, will be outside these limits in only $100\% \times \epsilon$ of the total observation time.

The traffic model presented here allows the formulation of a simple criterion for finding anomalous network states: if several consecutive measurements go beyond the confidence interval of $\epsilon=0.05$, we can confidently consider problems on the network. If we collect the data every 5 minutes, then the statistics of a few hours will make it possible to determine all the parameters of equation 1 with reasonable accuracy.

Presumably, the network state will be out of the confidence interval during the progress of a network attack. During port scanning, the number of active flows will increase with a nearly constant load, as the data transmitted is only limited to establish the connection and to close it. The channel load as well as the flow number should sharply increase during the progress of DDoS attacks. In order to prove these hypotheses it was decided to conduct two experiments with network scanning and with a DDoS attack.

## 3 Experiment

In order to clarify the details of unauthorized intrusion, it was decided to perform experiments that emulated attempted attacks. Experiments were carried out on the network of the Samara State Aerospace University (SSAU).

Remote machines were used as the source of the attack which were located in an external network. The utility Nmap was applied for port scanning, which was ordered to carry out a full scan of all hosts on the network.

A Web server was selected as the target during the progress of the DDoS attack. A few computers located in the external network were the sources of the attack. In the first part of the experiment the attacking computers sent ping requests simultaneously within half an hour. In the second part of the experiment the target computers were attacked (DDoS attack) with the help of a specialised program, LOIC. The Web server was attacked with the use of different types of traffic (HTTP, UDP, TCP) over an hour. Data were collected from all experiments, which are then analysed to identify patterns of different types of attacks.

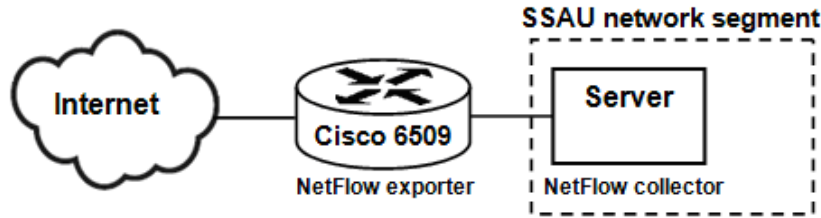

**Fig. 1.** The experimental scheme

Flow data, that are the basis for the analysis, were collected from the boundary router Cisco 6509 of the SSAU network. NetFlow collector nfdump [15] was used to gather data from the router. NetFlow export data is taken for analysis at regular intervals of five minutes. A file with the parameters of all the flows recorded on the router is formed every five minutes. The parameters are listed in the introduction, and include the beginning of the stream, the duration of the stream, the data transfer protocol, source address and port, destination address and port, the number of transmitted packets, and the amount of transferred data in bytes.

The analysis of data collected during network scanning has revealed a sharp increase in the number of active flows for almost the same amount of traffic transferred (see Figure 2). Each scanning computer generated of the order of 10-20 thousand of very short flows (up to 50 bytes) within 5 minutes. In the testing period the total number of active flows on a router that is generated by all sources is about 50-60 thousand.

Figure 2 shows a graph of the network states, the X-axis displays the number of completed flows N, the Y-axis displays the total load in Megabits per second (Mbps). Each point on the graph reflects the network state of the preceding five-minute interval, showing the dependence of the average channel load on the number of active flows. The points correspond to the normal network state and the triangles describe the state of the network, registered during a port scanning. Segments are depicted on the graph's parallel vertical axis and show the confidence intervals for the average load calculated for five flow intervals (20000-30000, 30000-40000, 40000-50000, 50000-60000, 60000-70000).

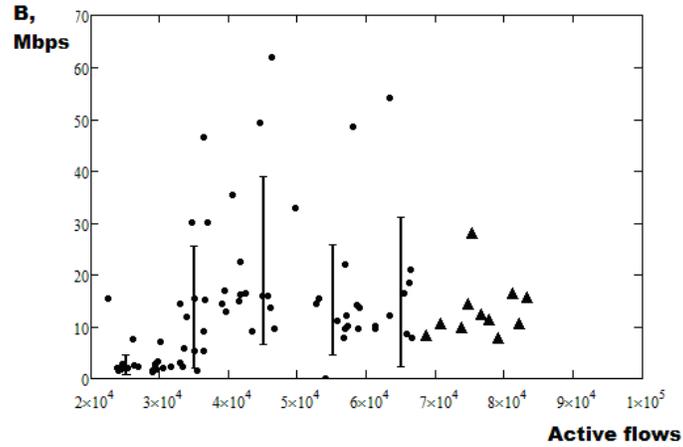

**Fig. 2.** Port scanning

As a result of the experiment with the ping requests, it was found that every attacking computer accounts for a very long flow of ICMP traffic, if we send requests through a single port. The data has been subsequently written into a nfdump file after the attack is finished, making it difficult to detect. It should be noted that one active ICMP flow to identify the occurrence of a failure in the information system is clearly inadequate; the number must extend to the tens of thousands of requests.

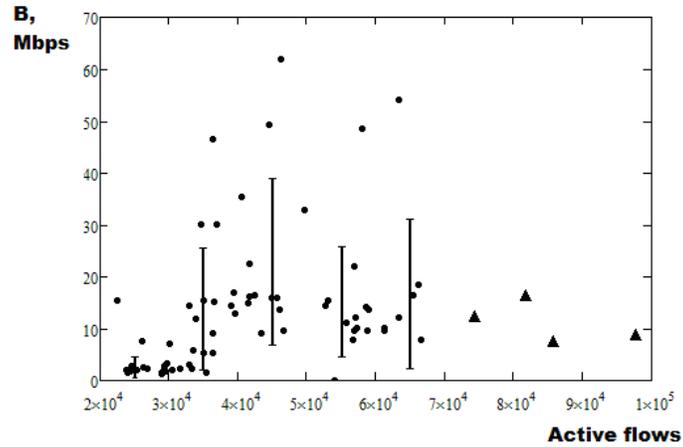

**Fig. 3.** DDoS attack

The analysis of modelling the DDoS attack by the LOIC utility also showed a sharp increase in the number of active flows, along with an increase in the traffic. The utility sends data in parallel to different ports of the target, thereby creating a large number of short flows for up to a minute (see Figure 3). The triangles show the network states recorded during the attack.

Thus, it becomes apparent that the NetFlow protocol may not only reveal the start of the attack, but also determine its type. A detailed description of attack detecting algorithms and work to create secure hosting services may be found in the following sections.

## 4   Algotithm for intrusion detection

Our studies have revealed patterns, based on the NetFlow data, that allow the IP addresses of the computers with which conducted DDoS attacks and port scans to be determined. Based on these patterns we developed an algorithm for attack detection. Before formulating the algorithm we will specify the format for recording flow data:

- Date and time of flow
- Duration of flow (in seconds, up to thousandths)
- Transfer protocol
- IP address and source port
- IP address and destination port
- The number of transmitted packets
- The number of bytes transferred

The algorithm developed for the detection of attacks such as DDoS and port scanning is:

1. Find IP-addresses of sources that generate a large number of flows.
   (a) If the size of these flows is very short, up to 50 bytes, it is most likely port scanning.
   (b) If the duration of the flows is greater then this IP-address might be carrying out DoS attacks.
2. Find IP-addresses of sources that generate very long streams (lasting more than 5 minutes). The IP-address assignment can be carried out DoS attack in this case.

If many IP-addresses from which a potential DoS attack are found, we may classify this attack as DDoS.

In order to prevent any network attack, early detection is important to enable steps to be taken to neutralise it. NetFlow data comes from the router from time to time, depending on the settings. At the same time, a balance between the frequency of collection of flow statistics and the time needed for processing is also needed. Therefore it was decided to establish the frequency of querying the NetFlow data to once a minute.

It should be noted that the NetFlow statistics provide information on flows that are already completed. Since the flow is considered as active for a certain time after its completion, completed flows also need to be considered active.

## 5   DDoS attacks study centre (secure hosting creation)

We have developed practical algorithms that are implemented as a script in Perl. The script has been installed on the protected server. NetFlow data comes to the server running the NetFlow nfdump collector from the boundary router (BGP) on the SSAU network every minute. The script receives a file with data on entering flows. These data are processed by a script in accordance with the attack detecting algorithm described in the previous section. A list of suspicious IP addresses, from which an attack may be carried out, are produced as the output of the script.

The processing time of the NetFlow data is very small (tens of a millisecond), whereas the intrusion detection addresses will be equal to the period of the export data from the router, i.e. one minute.

Suspicious IP addresses are entered into the database and all traffic from those addresses are blocked by an iptables firewall [16] for 5 minutes. Iptables is installed on the protected server, i.e. only the server is protected, not the whole network. If necessary, the protection can be extended to a whole SSAU network, blocking suspicious IP address on the boundary router. In the coming year we plan to explore the possibilities of using the NetFlow protocol for the detection of DDoS attacks, for a combination of several basic types of attacks.

The problem of preventing DDoS attacks, as well as unauthorized network intrusion; do not lose their sharpness, so SSAU created the Centre for the Study of Network Attacks. The main purpose of the new centre is to develop new techniques to detect and prevent various types of unauthorized network intrusion. Hosting that is protected from DDoS attack has been created inside one segment of the university network. The method of protection is based on the method presented in this article. The server that is running the NetFlow collector receives NetFlow data from the boundary router of university network. This data is then processed to produce a "black list" of addresses that are blocked by an iptables firewall.

## 6   Conclusion

In this paper, the detection of attacks such as DDoS and port scans using a flow traffic model was proposed, based on receiving data according to the Cisco NetFlow protocol from the border routers. An experiment to test this model and create prevention algorithms has been described. The experimental results have confirmed that the proposed flow traffic model can be used effectively to detect these attacks.

An algorithm for detecting suspicious IP addresses that can go attack was suggested. These addresses can be used in intrusion prevention systems in order to block them. Also, the algorithm for the detection of suspicious addresses was implemented as a script that works in conjunction with a firewall iptables. This system of detecting and preventing attacks such as DDoS and port scanning was installed on the SSAU host network. In the future we plan to continue studying the possibility of using the NetFlow protocol to detect various types of unauthorized network intrusion. It is also planned to create a network protection system directly on the SSAU network boundary router using Cisco IOS features.